\documentclass[aps,prb,twocolumn,showpacs]{revtex4-1}
\usepackage{graphicx}
\usepackage{latexsym}
\usepackage{amssymb}
\usepackage{amsmath}
\usepackage{amsfonts}
\usepackage{bm}
\usepackage{multirow}
\usepackage{color}

\begin{document}

\title{Natural orbitals renormalization group approach to a Kondo singlet}

\author{Ru Zheng}
\author{Rong-Qiang He}
\email{rqhe@ruc.edu.cn}
\address{Department of Physics, Renmin University of China, Beijing 100872, China}
\author{Zhong-Yi Lu}
\email{zlu@ruc.edu.cn}
\address{Department of Physics, Renmin University of China, Beijing 100872, China}
\date{\today}

\begin{abstract}

A magnetic impurity embedded in a metal host is collectively screened by a cloud of conduction electrons to form a Kondo singlet below a characteristic energy scale $T_K$, the Kondo temperature, through the mechanism of the Kondo effect. The cloud of conduction electrons, named the Kondo screening cloud, is considered to spread out in real space over the so-called Kondo length $\xi_k={\hbar v_F}/{k_B T_K}$ with $v_F$ being the Fermi velocity, which however has not been detected experimentally even though $\xi_k$ is estimated as large as 1 $\mu$m. We have reinvestigated the Kondo singlet by means of the newly developed natural orbitals renormalization group (NORG) method.
We find that, in the framework of natural orbitals formalism, the Kondo screening mechanism becomes transparent and simple, while the intrinsic structure of Kondo singlet is clearly resolved. For a single impurity Kondo system, there exits a single active natural orbital which screens the magnetic impurity dominantly. In the perspective of entanglement, the magnetic impurity is entangled dominantly with the active natural orbital, i.e., the subsystem formed by the active natural orbital and the magnetic impurity basically disentangles from the remaining system. We have also studied the structures of the active natural orbital respectively projected into real space and momentum space. Moreover, the dynamical properties, represented by one-particle Green's functions defined at impurity site with active natural orbital, were obtained by using correction vector method. In order to clarify the spatial extension of the Kondo screening cloud, the concept of Kondo correlation energy was introduced. With this concept we obtain a characteristic length scale beyond which the Kondo screening cloud is hardly detected in experiment. Our numerical results indicate that this characteristic length scale usually is just a few nanometers, which interprets why it is difficult to detect the Kondo screening cloud experimentally in a metal host.
\end{abstract}


\date{\today} \maketitle

\section{Introduction}

The Kondo effect \cite{Kondo1964,Hewson1997} is one of the intensively studied problems in condensed matter physics, both theoretically and experimentally. It results from the
antiferromagnetic exchange interaction between a magnetic impurity and conduction electrons in a metal host or in a quantum dot. Below a characteristic energy scale $T_K$, namely the Kondo
temperature, the magnetic impurity is collectively screened by the surrounding conduction electrons, forming the Kondo screening cloud centered at the magnetic impurity with spatial
extension determined by the Kondo length \cite{Affleck1996,Affleck2001,Affleck2005,Holzner2009,Busser2010,Bergmann2007}, which is conventionally considered as $\xi_k={\hbar v_F}/{k_B T_K}$ with $v_F$ being the Fermi velocity. Typically, the Kondo temperature $T_K$ is about 1 K and the Fermi velocity $v_F$ is about $10^5 \sim 10^6$ m/s, thus the Kondo length $\xi_k$ is estimated to be 1 $\mu$m, which is about thousands times of lattice constant.

The standard single-impurity Kondo model and single-impurity Anderson model are two well known models to represent quantum impurity problems. The ground state of these models is a collective spin singlet, called the Kondo singlet, formed by the magnetic impurity and conduction electrons. The ground state wave function extends in the real space over a region of length scale $\xi_k$, called the Kondo screening cloud. All the conduction electrons within the screening cloud involve to screen the magnetic impurity spin, and the screened complex acts like a potential scatter for the electrons outside the screening cloud\cite{Affleck1996}, with a $\pi/2$ phase shift at the Fermi energy. The screening mechanism of Kondo effect has been studied extensively, however, it remains elusive when it comes to the problem of the spatial features of the Kondo screening cloud. Many proposals \cite{Affleck2001,Gubernatis1987,Barzykin1996,Pascal2003,Hand2006,Affleck2008,Pereira2008,Jinhong2013,Borda2007,Yoshii2011,Simon2002} for the measurement of Kondo screening cloud
have been put forward, however, it has never been detected in experiment to support the existence of the Kondo screening cloud\cite{Boyce1974,Madhavan1998,Manoharan2000,Wenderoth2011,Y2007}, even though the Kondo length is considered as large as 1 $\mu$m.

Theoretically, various approaches have been developed and extensively used to study the Kondo effect, among which the numerical renormalization group (NRG) \cite{Wilson1975, Bulla2008}, the Bethe
ansatz \cite{Andrei1983, Tsvelik1983}, and the density matrix renormalization group (DMRG) \cite{White1992, White1993, U.S.2005} are accurate ones. Nevertheless, the spatial structure and properties of the
Kondo cloud are difficult to obtain by using the powerful NRG and Bethe ansatz methods, since these two approaches work in the moment space. In contrast, the DMRG preserves the detailed information of lattice geometry in real space, but it is efficient mainly in 1D systems rather than 2D and 3D systems, even though there has been progress
toward the techniques of DMRG \cite{B¨¹sser2013,Shirakawa2014,Shirakawa2016}. For both NRG and DMRG, the number of impurities or channels in a quantum impurity system is very restricted, generally to single impurity or single channel. In this work, we used a newly developed numerical approach, namely the natural orbitals renornalization group (NORG)\cite{He2014}, which preserves the whole geometric details of lattice to study the Kondo screening problem. More importantly, the intrinsic structure of a Kondo singlet can be clearly resolved in the framework of natural orbitals formalism. A rather simple and transparent picture of Kondo screening mechanism is thus obtained.

This paper is organized as follows. In Sec.~\ref{sec:Model-Method} the models (\ref{sec:SIK-Model}) and NORG numerical method (\ref{sec:Numerical method}) are introduced. In Sec.~\ref{sec:Results}, the natural orbitals occupancy (\ref{sec:Active natiral orbitals}) is first examined, and the number of active natural orbitals (ANo) of the system is confirmed as only one. We then demonstrate the Kondo Screening mechanism within natural orbitals formulation (\ref{sec:Kondo Screening mechanism}), i.e., in the view of active natural orbital, the spin-spin correlation function and entanglement entropy are presented. The structures of the active natural orbital respectively projected into real space and momentum space are further presented in Sec.~\ref{sec:Amplitudes}. The dynamical properties, represented by the one-particle Green's function, are presented in Sec.~\ref{sec:Spectral function}. We examine the spatial extension of the Kondo screening cloud in Sec.~\ref{sec:Screening cloud} and then interpret why it is difficult to detect experimentally in a metal host. Based on the concept of Kondo correlation energy, we further obtain a characteristic length scale beyond which the Kondo screening cloud is hardly detected in experiment. Section~\ref{sec:summary} gives the discussion with the above studies as well as a short summary of this work.

\section{Modeles and numerical method}
\label{sec:Model-Method}
\subsection{Models}
\label{sec:SIK-Model}
We studied both single-impurity Kondo model and single-impurity Anderson model, each of which is divided into two parts namely magnetic impurity and bath. These two models can be transformed into each other by the Schrieffer-Wolff transformation. We modeled the bath part by a non-interacting tight-binding band on a lattice with arbitrary dimension and geometry, represented by the following Eq.~(\ref{eq:Hband}), meanwhile we set the magnetic impurity coupled to the central site of the bath part,
\begin{equation}
H_{\text{band}}=-t\sum\limits_{\langle ij\rangle\sigma}(c_{i\sigma }^ {\dagger}{c_{j\sigma }}+ h.c.),
\label{eq:Hband}
\end{equation}
where $c_{i\sigma }^ {\dagger}({c_{i\sigma }})$ denotes the creation (annihilation) operator of a conduction electron at the $i$-th site with spin component $\sigma=\uparrow,\downarrow$, and $t$ is the nearest-neighbor hopping integral of the tight-binding band.

\begin{figure}[h!]
\centering
\includegraphics[width=7.0cm]{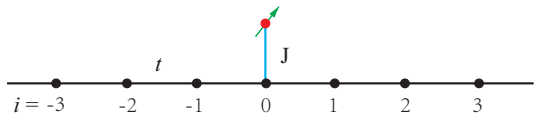}
\caption{\label{fig:Model}(color online) Schematic of the single-impurity Kondo or Anderson model on a chain. Here $i$ denotes the index of lattice site and the impurity marked by the red dot is coupled directly to the central site indexed as site 0 in the chain.}
\end{figure}

The total Hamiltonian of single-impurity Kondo model on a lattice reads
\begin{equation}
\begin{array}{l}
H = H_{\text{band}} + H_\text{Kondo};\\
H_{\text{Kondo}}=J{\bf{S}}_{\text{imp}} \cdot {\bf{s}}({r_i}).
\end{array}
\label{eq:Kondo}
\end{equation}
Here $H_{\text{Kondo}}$ describes the Kondo interaction between a magnetic impurity with spin ${\bf{S}}_{\text{imp}}$ and the conduction electron spin ${\bf{s}}( {r_i})=\frac{1}{2}\sum_{\alpha\beta}c_{i\alpha}^ {\dagger}{{\bf{\sigma}}_{\alpha\beta}}{c_{i\beta}}$ located at position $r_i$ with antiferromagnetic coupling $J>0$, where ${\bf{\sigma}}$ represents the vector of Pauli matrices. Figure~\ref{fig:Model} is a schematic view of the single-impurity Kondo model on a chain, in which the magnetic impurity is coupled to the central site indexed as site 0 in the chain. 

Then we consider the total Hamiltonian of single-impurity Anderson model as follows,
\begin{equation}
\begin{array}{l}
H = H_{\text{band}} + H_\text{imp} + H_\text{hyb};\\
H_{\text{imp}}=-\mu(n_{\text{imp},\uparrow} + n_{\text{imp},\downarrow}) + Un_{\text{imp},\uparrow}n_{\text{imp},\downarrow};\\
H_{\text{hyb}}=V\sum\limits_\sigma(c_{i\sigma}^ {\dagger}c_{\text{imp},\sigma}+ c_{\text{imp},\sigma }^ {\dagger}c_{i\sigma}).
\end{array}
\label{eq:Anderson}
\end{equation}
Here, $n_{\text{imp},\uparrow}(n_{\text{imp},\downarrow})$ denotes the occupacy number operator with spin component $\sigma=\uparrow(\downarrow)$ acting on the impurity site, $\mu$ the single-particle energy of the impurity, $U$ the strength of Hubbard interaction at the impurity site, $c_{\text{imp},\sigma }^ {\dagger}(c_{\text{imp},\sigma})$ the creation (annihilation) operator at the impurity site with spin component $\sigma=\uparrow,\downarrow$, and $V$ the hybridization between the impurity site and the conduction electrons.

The single-impurity Kondo model is employed in all the following Sections except Sec.~\ref{sec:Spectral function} in which the single-impurity Anderson model is employed. Throughout the whole work, the tight-binding band with an odd number of sites was adopted. That is, the overall number of sites in the whole system is even. In the calculations, we set the
nearest-neighbor hopping integral $t=1$ and kept half-filling of the conduction band. For the Anderson model, $\mu$ was set to $U/2$, which keeps the impurity Hamiltonian $H_{\text{imp}}$ particle-hole symmetric and the chemical potential of the whole system being zero. Periodic boundary condition was used. All the calculations were carried out in the subspace with $S_{\rm total}^z=0$.

\subsection{Numerical method}
\label{sec:Numerical method}
We adopted a newly developed numerical many-body approach, namely the natural orbitals renormalization group (NORG) \cite{He2014}, to study the screening mechanism of Kondo effect. The NORG method works efficiently on quantum impurity models in the whole coupling regime, and it preserves the whole geometric information of a lattice. We emphasize that the effectiveness of the NORG is independent of any topological structure of a lattice.\cite{He2014} And the NORG has recently been applied to solve the well-known two-impurity Kondo critical point issue. \cite{he2015}

Consider an interacting N-electron correlated system with an orthonormal set of one-electron states $|i\sigma\rangle=c_{i\sigma }^ {\dagger}|{\text{vac}}\rangle$ at the $i$-th site, where $|{\text{vac}}\rangle$ is the vacuum state. Then for a normalized many-body wave function $|\Psi\rangle$ of the system, the single-particle density matrix $D$ can be defined by its elements $D_{ij}  = \sum_{\sigma}\langle\Psi| c_{i\sigma}^{\dagger}c_{j\sigma}|\Psi\rangle$. The natural orbitals and their occupancy numbers correspond to the complete set of eigenvectors and eigenvalues of the single-particle density matrix $D$. A natural orbital is called an active natural orbital if its occupancy is about 1, otherwise an inactive orbital. In Ref.~\onlinecite{He2014}, it has been shown that the number of the Slater determinants in the expansion of $|\Psi\rangle$ is determined by the number of active natural orbitals with inactive natural orbitals frozen as background, while the number of active natural orbitals is roughly equal to the number of quantum impurities for a quantum impurity system.

The NORG method works in the Hilbert space constructed from a set of natural orbitals\cite{He2014}. Its realization essentially involves a representation transformation from site representation into natural orbitals representation through iterative orbital rotations. More specifically, one performs the representation transformation from site representation into natural orbitals representation by $|m\sigma\rangle=\sum_{i=1}^NU_{mi}^ {\dagger}|i\sigma\rangle$, where $|m\sigma\rangle=d_{m\sigma }^{\dagger}|{\text{vac}}\rangle$ is a one-particle state at the $m$-th orbital in the natural orbitals representation with $d_{m\sigma }^ {\dagger}$ being the corresponding creator, and $U$ is an $N \times N$ unitary matrix. Then the creation operators can be transformed from site representation into natural orbitals representation by $d_{m\sigma }^ {\dagger}=\sum_{i=1}^NU_{mi}^{\dagger}c_{i\sigma }^ {\dagger}$, and the corresponding inverse transformation is realized by $c_{i\sigma }^ {\dagger}=\sum_{m=1}^NU_{im}d_{m\sigma }^ {\dagger}$.

In practice, to efficiently realize the NORG, we only rotate the orbitals of bath, that is, only the orbitals of bath are transformed into natural orbitals representation. After the NORG representation transformation, the Kondo interaction  $H_{\text{Kondo}}$ in the framework of natural orbitals formalism is given by
\begin{equation}
\begin{split}
H_{\text{Kondo}}=&\frac{J}{2}\sum\limits_{mn}U_{im}U_{in}^*S_{\text{imp}}^z(d_{m\uparrow }^ {\dagger}d_{n\uparrow }-d_{m\downarrow}^{\dagger}d_{n\downarrow}) \\
                 & -\frac{J}{2}\sum\limits_{mn}U_{im}U_{in}^*c_{\text{imp},\uparrow }^ {\dagger}d_{n\uparrow }d_{m\downarrow }^ {\dagger}c_{\text{imp},\downarrow }  \\
                 & -\frac{J}{2}\sum\limits_{mn}U_{im}U_{in}^*d_{m\uparrow }^ {\dagger}c_{\text{imp},\uparrow }c_{\text{imp},\downarrow }^ {\dagger}d_{n\downarrow }
\label{eq:HamiltonianKondo}
\end{split}
\end{equation}
with $c_{\text{imp},\uparrow }^ {\dagger}(c_{\text{imp},\uparrow })$ denoting the creation (annihilation) operator with spin up at the impurity site.

\section{Numerical results}
\label{sec:Results}
\subsection{Active natural orbitals}
\label{sec:Active natiral orbitals}
We first examine the natural orbitals occupancies. As shown in Ref.~\onlinecite{He2014}, for a quantum impurity system most of the natural orbitals exponentially rush into doubly
occupancy or empty. By comparison, only a small number of natural orbitals deviate well from full occupancy and empty, namely active natural orbitals, which play a substantial role in constructing the ground state wave function. The number of active natural orbitals is about the number of interacting impurities for a quantum impurity model. This is the underlying basis for the NORG working on quantum impurity systems.

\begin{figure}[h!]
\centering
\includegraphics[width=\columnwidth]{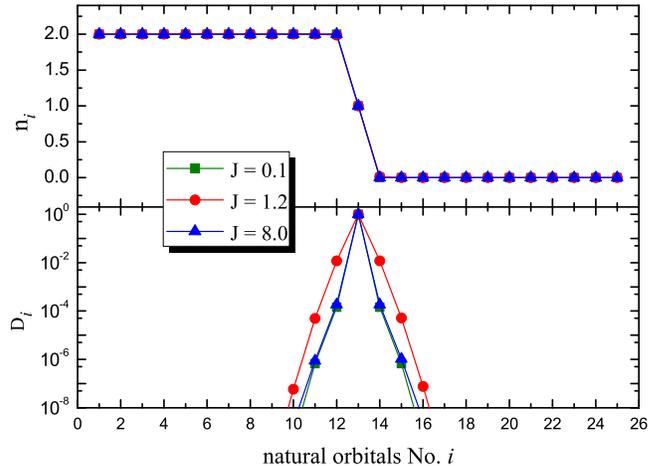}
\caption{\label{fig:NOOcc}(color online) Calculated natural orbitals occupancies $ n_i $ and their corresponding deviations $D_i={\text{min}}(n_i,2-n_i)$ from full occupancy or empty for the ground state of the single-impurity Kondo model (Eq.~(\ref{eq:Kondo})) with  Kondo coupling $J $=0.1, 1.2, and 8.0 respectively. Here all the natural orbitals exponentially rush into full occupancy or empty, except the active natural orbital, which is about half-occupied. The calculation was carried out for a chain with 25 lattice sites.}
\end{figure}

Figure~\ref{fig:NOOcc} shows the calculated natural orbitals occupancies $ n_i $ for the ground state of the single-impurity Kondo model represented by Eq.~(\ref{eq:Kondo}). The corresponding deviations of natural orbitals occupancies from full occupancy or empty are also shown in Fig.~\ref{fig:NOOcc}, which are given by $D_i={\text{min}}(n_i,2-n_i)$. As we see, all the natural orbitals exponentially rush into full occupancy or empty except for only one natural orbital with half-occupancy, namely the active natural orbital. In the following sections, we investigate the screening mechanism of Kondo effect in the natural orbitals representation, especially in the view of active natural orbital.

\subsection{Kondo screening mechanism}
\label{sec:Kondo Screening mechanism}
In this section, we study the screening mechanism of single-impurity Kondo problem represented by Eq.~(\ref{eq:Kondo}) in the natural orbitals representation. To this end, we calculated the spin-spin correlation function between the active natural orbital (ANo) and magnetic impurity $\langle {\bf{S}}_{\text{imp}}^z{\bf{s}}_{\text{ANo}}^z\rangle$ as well as the von Neumann entropy of the system as functions of the Kondo coupling $J$. Here ${\bf{s}}_{\text{ANo}}$ stands for the spin of electron occupying the active natural orbital and $\langle {\bf{S}}_{\text{imp}}^x{\bf{s}}_{\text{ANo}}^x\rangle = \langle {\bf{S}}_{\text{imp}}^y {\bf{s}}_{\text{ANo}}^y\rangle = \langle {\bf{S}}_{\text{imp}}^z{\bf{s}}_{\text{ANo}}^z\rangle$ due to the isotropy of the Kondo term in Eq.~(\ref{eq:Kondo}). For simplicity, we performed the calculations on a 1D chain, schematically shown in Fig.~\ref{fig:Model}.

\begin{figure}[h!]
\centering
\includegraphics[width=\columnwidth]{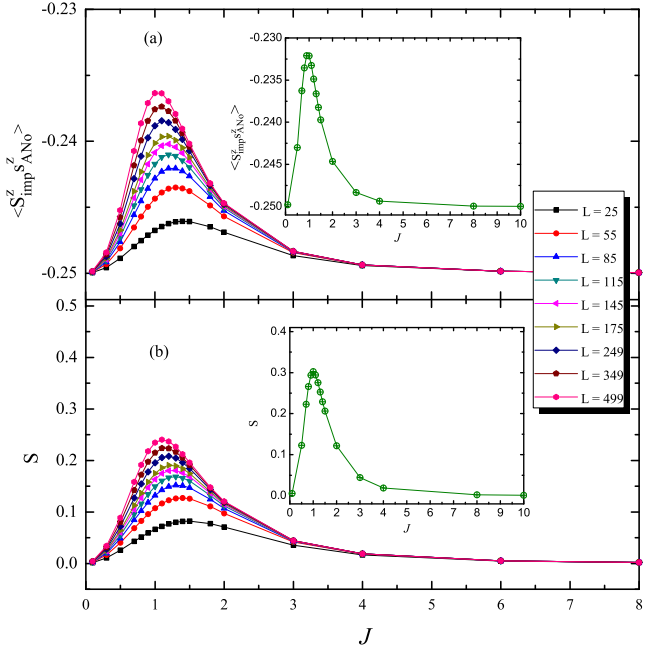}
\caption{\label{fig:KSMc-FS-TL}(color online) Numerical results calculated on a 1D chain: (a) spin-spin correlation function
$\langle{\bf{S}}_{\text{imp}}^z{\bf{s}}_{\text{ANo}}^z\rangle$ between the active natural orbital and the magnetic impurity and (b) von Neumann entanglement entropy $S$ between the subsystem A and B, versus the Kondo coupling $J$. The corresponding insets show the values in the thermodynamic limit $L\to\infty$.}
\end{figure}

\begin{figure}[h!]
\centering
\includegraphics[width=\columnwidth]{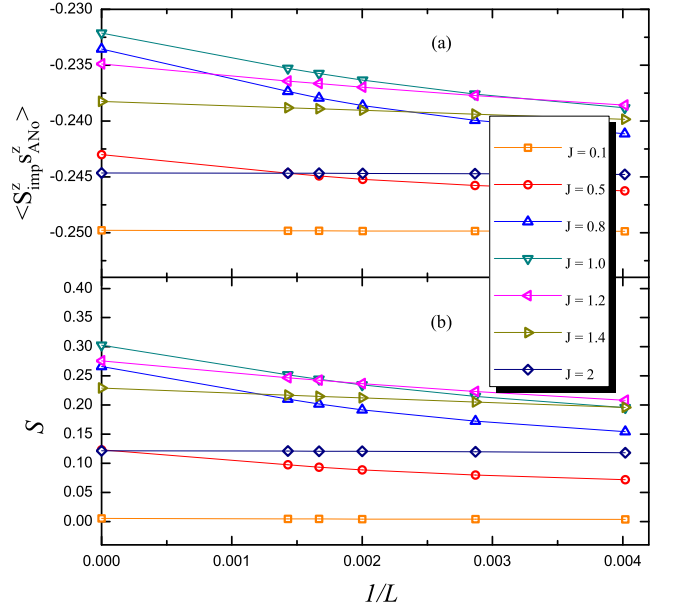}
\caption{\label{fig:Thermodynamic1}(color online) The finite size extrapolation of (a) spin-spin correlation function
$\langle{\bf{S}}_{\text{imp}}^z{\bf{s}}_{\text{ANo}}^z\rangle$ between the active natural orbital and the magnetic impurity and (b) von Neumann entanglement entropy $S$ between the subsystem A and B calculated on a 1D chain. The endpoints at ${1}/{L} \to 0$ in the curves are obtained by a quadratic polynomial fit of the five points with $L = 249,349,499,599$ and $699$.}
\end{figure}

\begin{figure}[h!]
\centering
\includegraphics[width=\columnwidth]{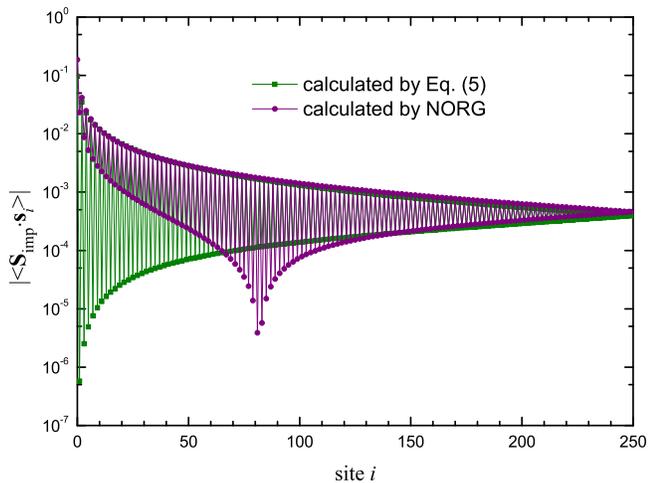}
\caption{\label{fig:Spin-Fun}(color online) Spin-spin correlation function $\langle {\bf{S}}_{\text{imp}} \cdot {\bf{s}}_i \rangle$ between the magnetic impurity and site $i$ calculated on a chain of 499 lattice sites respectively by Eq.~(\ref{eq:wave-function}) and NORG method with the Kondo coupling $J=1.0$. The envelope of the spin-spin correlation function obtained by Eq.~(\ref{eq:wave-function}) is consistent with the results calculated by the NORG.}
\end{figure}

Numerical results calculated on the 1D chain with different sizes are plotted in Fig.~\ref{fig:KSMc-FS-TL}(a), and the corresponding inset shows the results in the thermodynamic limit $L\to\infty$. The values in the thermodynamic limit are obtained by finite size extrapolation with a quadratic polynomial fit, as shown in Fig.~\ref{fig:Thermodynamic1}(a). In both the weak coupling limit $J \to 0$ and the strong coupling limit $J \to \infty$, the spin-spin correlation function $\langle {\bf{S}}_{\text{imp}}^z{\bf{s}}_{\text{ANo}}^z\rangle \to -\frac{1}{4}$. This indicates that in both limits, the active natural orbital screens the impurity local moment solely. Therefore the active natural
orbital fully forms a spin singlet with the impurity spin ${\bf{S}}_{\text{imp}}$ in both limits, which results in a concise expression for the ground state wave function $|\Psi_0 \rangle$ as follows,
\begin{equation}
|\Psi_0 \rangle=\frac{1}{\sqrt 2}({|\uparrow \rangle}_{\text{imp}}d_{\text{ANo}\downarrow }^ {\dagger}-{|\downarrow \rangle}_{\text{imp}}d_{\text{ANo}\uparrow }^ {\dagger})\otimes|0\rangle,
\label{eq:wave-function}
\end{equation}
where $|0\rangle=\prod_{m=1}^{(L-1)/2} d_{m\uparrow }^ {\dagger}d_{m\downarrow }^ {\dagger}|{\text{vac}}\rangle$ and the index of the active natural orbital is $(L+1)/2$. Thus the spin singlet formed by the magnetic impurity and the active natural orbital disentangles from the Fermi sea $|0\rangle$ formed by the remaining natural orbitals.
In the intermediate regime of $J$, the spin-spin correlation function $\langle{\bf{S}}_{\text{imp}}^z {\bf{s}}_{\text{ANo}}^z\rangle \sim -\frac{1}{4}$ and the active natural orbital screens more than $90\%$ of the impurity spin, as shown in the inset of Fig.~\ref{fig:KSMc-FS-TL}(a), which indicates that the active natural orbital screens the impurity spin dominantly. In this case, the ground state energy computed by using the wave function Eq.~(\ref{eq:wave-function}) can be accurate within three or even up to five digits (not shown). Figure~\ref{fig:Spin-Fun} shows the spin-spin correlation function $\langle {\bf{S}}_{\text{imp}} \cdot {\bf{s}}_i \rangle$ between the magnetic impurity and site $i$ calculated on a chain of 499 lattice sites with the Kondo coupling $J=1.0$ by using Eq.~(\ref{eq:wave-function}) and the NORG approach respectively. As we see, the envelope of the spin-spin correlation function obtained by using Eq.~(\ref{eq:wave-function}) is consistent with the one obtained by using the NORG approach.

Accordingly, we arrive at such a physical understanding. There exits only one active natural orbital for the single impurity Kondo problem in the framework of natural orbitals formalism. In the whole Kondo coupling regime, the magnetic impurity spin is screened dominantly by the active natural orbital and the Eq.~(\ref{eq:wave-function}) is a valid approximation for studying the ground state.

To further enrich the understanding, we study the von Neumann entanglement entropy of the system in the framework of natural orbitals formalism. The whole system is divided into two subsystems A and B, in which subsystem A is formed by the magnetic impurity and the active natural orbital, while subsystem B is formed by the remaining natural orbitals. The von Neumann entanglement entropy $S$ between the two subsystems A and B is defined as
\begin{equation}
\begin{array}{l}
S=-{\rm Tr}(\rho_{\text{A}}\log_2\rho_{\text{A}})=-{\rm Tr}(\rho_{\text{B}}\log_2\rho_{\text{B}}),
\end{array}
\end{equation}
where $\rho_{\text{A}} (\rho_{\text{B}})$ is the reduced density matrix of the subsystem A (B), $\rho_{\text{A}}={\rm Tr}_{\text{B}}\rho_{\text{AB}}$ and $\rho_{\text{B}}={\rm Tr}_{\text{A}}\rho_{\text{AB}}$ with $\rho_{\text{AB}}$ being the density matrix of the whole system AB. Noting that the number of the degrees of freedom is 4 for the active natural orbital and 2 for the magnetic impurity, the number of the degrees of freedom is 8 for $\rho_{\text{A}}$. Thus, the von Neumann entanglement entropy is 0 in the case of subsystem A being disentangled from B, while it is equal to 3 if subsystem A and B are maximally entangled. The calculated results are plotted in Fig.~\ref{fig:KSMc-FS-TL}(b) and the inset presents the values of the entropy $S$ in the thermodynamic limit $1/L \to 0$, which are obtained by finite size extrapolation with a quadratic polynomial fit, as shown in Fig.~\ref{fig:Thermodynamic1}(b). In addition, we notice that a similar work was done recently \cite{Adrian E. Feiguin 2017}.

From Figs.~\ref{fig:KSMc-FS-TL} and~\ref{fig:Thermodynamic1},we find that the entropy $S\to 0$ in both the weak Kondo coupling $J\to 0$ and strong Kondo coupling $J\to \infty$ regimes, while in the intermediate Kondo coupling regime the entropy $S$ is finite (about 0.3) but very small in comparision with the saturation value 3. The calculated results thus show that the subsystem A disentangles from the subsystem B in both the weak and the strong Kondo coupling regimes, while the subsystem A entangles weakly with the subsystem B in the intermediate Kondo coupling regime. This also indicates that the magnetic impurity entangles with the active natural orbital solely in both the weak and the strong Kondo coupling regimes, while the magnetic impurity entangles dominantly with the active natural orbital in the intermediate Kondo coupling regime. As we have seen, the results of the von Neumann entanglement entropy are consistent with the results of spin-spin correlation function $\langle{\bf{S}}_{\text{imp}}^z {\bf{s}}_{\text{ANo}}^z\rangle$.

\begin{figure}[h!]
\centering
\includegraphics[width=\columnwidth]{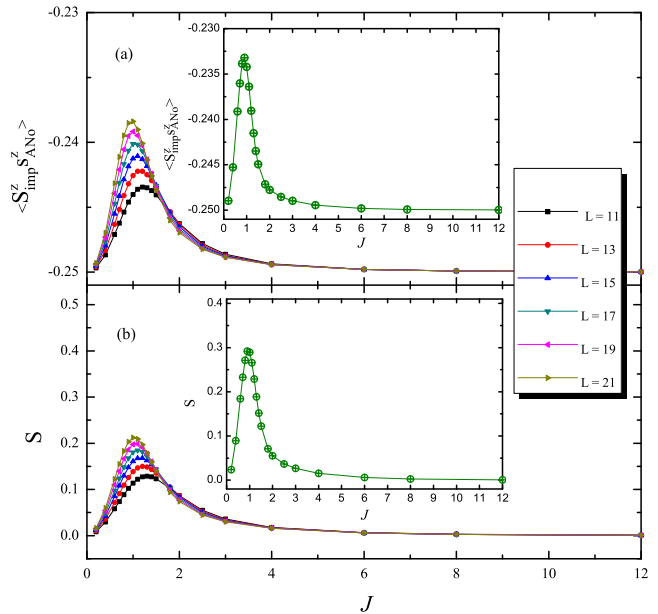}
\caption{\label{fig:KSMsl-FS-TL}(color online) Numerical results calculated on a square lattice: (a) spin-spin correlation $\langle {\bf{S}}_{\text{imp}}^z{\bf{s}}_{\text{ANo}}^z\rangle $ between active natural orbital and the magnetic impurity and (b) von Neumann entanglement entropy $S$ between the subsystem A and B, as functions of the Kondo coupling $J$. The corresponding insets show the values in the thermodynamic limit $L^2\to\infty$.}
\end{figure}

\begin{figure}[h!]
\centering
\includegraphics[width=\columnwidth]{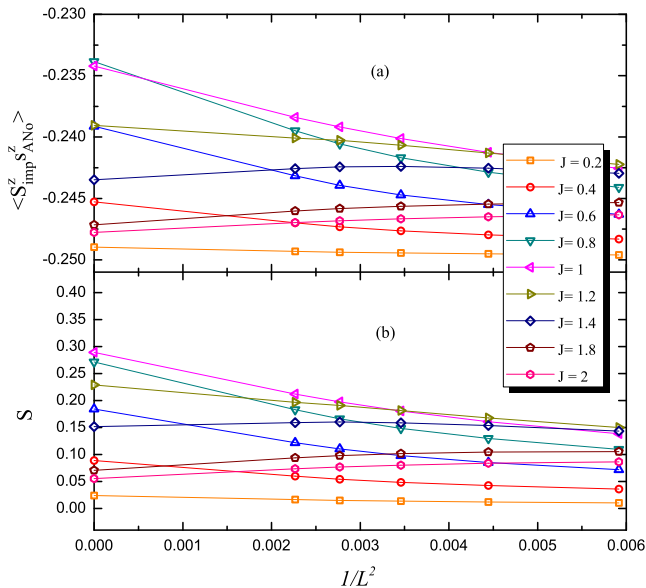}
\caption{\label{fig:Thermodynamic2}(color online) The finite size extrapolation of (a) spin-spin correlation function $\langle{\bf{S}}_{\text{imp}}^z{\bf{s}}_{\text{ANo}}^z\rangle$ between active natural orbital and impurity and (b) von Neumann entanglement entropy between the subsystem A and B calculated on a square lattice. The endpoints at ${1}/{L^2} \to 0$ in the curves are obtained by a quadratic polynomial fit of the five points with $L = 13,15,17,19$ and $21$.}
\end{figure}

We further performed the corresponding calculations on a square lattice. Numerical results calculated on the square lattice with finite size $L\times L$ ($L$ is odd) are plotted in Fig.~\ref{fig:KSMsl-FS-TL} and then the corresponding values in the thermodynamic limit are also obtained by finite size extrapolation with a quadratic polynomial fit, as presented in Fig.~\ref{fig:Thermodynamic2}. Similar to the case of the 1D chain, there is also only one active natural orbital for the ground state of the single-impurity Kondo model Eq.~(\ref{eq:Kondo}) on  the square lattice and the active natural orbital screens the magnetic impurity moment dominantly. Actually, we have done the calculations on many different lattices and arrive at the same physical understanding. Thus, the screening mechanism of single impurity Kondo problem with only one active natural orbital is universal and independent of the topology or geometry of the lattice.

\subsection{Structures of the active natural orbital}
\label{sec:Amplitudes}
To clarify the structures of the active natural orbital, we examine the active natural orbital in real space (site representation, namely Wannier representation) and momentum space respectively. From the representation transformation $|m\sigma\rangle=\sum_{i=1}^NU_{mi}^ {\dagger}|i\sigma\rangle$ used in the NORG, we can easily obtain the amplitude of the $m$-th natural orbital projected into the $i$-th Wannier orbital in real space, namely $|w_i^m|^2=|U_{mi}^ {\dagger}|^2$. The amplitude of the $m$-th natural orbital $|u_k^m|^2$ projected into a Bloch state with wavevector $k$ in momentum space, via $d_{m\sigma }^ {\dagger}=\sum_{k}u_{k}^mc_{k\sigma }^ {\dagger}$, can be obtained by the Fourier transform of operators $c_{l\sigma }^ {\dagger}=\frac{1}{\sqrt{N}}\sum_{k}e^{-ikl}c_{k\sigma }^ {\dagger}$, namely $u_k^m=\frac{1}{\sqrt{N}}\sum_{l=1}^Ne^{-ikl}U_{ml}^ {\dagger}$. On the other hand, the decomposition of a Bloch state with wavevector $k$ into the natural orbitals can be realized by the Fourier transform of operators $c_{k\sigma }^ {\dagger}=\frac{1}{\sqrt{N}}\sum_{l=1}^Ne^{ikl}c_{l\sigma }^ {\dagger}$ in combination with the NORG inverse transformation $c_{i\sigma }^ {\dagger}=\sum_{m=1}^NU_{im}d_{m\sigma }^ {\dagger}$. The amplitude of a Bloch state with wavevector $k$ projected into the $m$-th natural orbital $|\beta_{m}^k|^2$ can thus be obtained by $c_{k\sigma }^ {\dagger}=\sum_{m=1}^N\beta _{m}^k d_{m\sigma }^ {\dagger}$, namely $\beta _{m}^k=\frac{1}{\sqrt{N}}\sum_{l=1}^Ne^{ikl}U_{lm}$. For the occupancy number of electrons $n^k$ in a Bloch state with wavevector $k$, it is obtained by $n^k=\sum_{m=1}^Nn_{m}|\beta_{m}^k|^2$ with $n_{m}$ being the occupancy number of the $m$-th natural orbital. Thus, the occupancy number of electrons $n(\varepsilon_k)$ at energy level $\varepsilon_k$ can be obtained. To inspect the effect of the Kondo coupling, it is more meaningful to compute the variation of electron occupancy number $\Delta n(\varepsilon_k)$ with respect to the case of the Kondo coupling $J$=0, in which the electrons obey the Fermi distribution in momentum space. All the numerical results in this subsection were calculated for the ground state of the single-impurity Kondo model (Eq.~(\ref{eq:Kondo})) on a chain with 499 lattice sites.

\begin{figure}[h!]
\centering
\includegraphics[width=1\columnwidth]{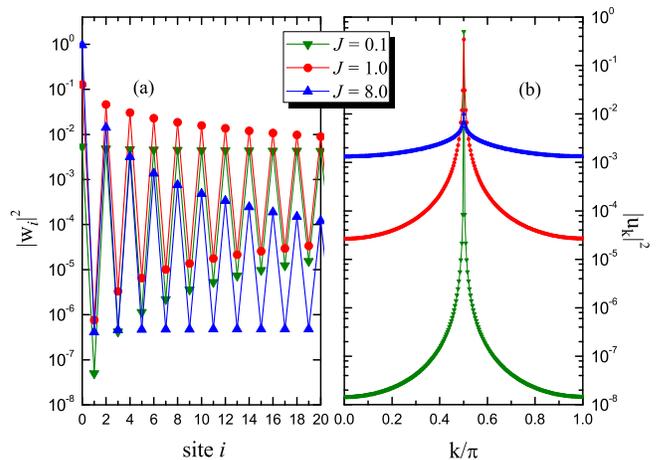}
\caption{\label{fig:NOAmp}(color online) Amplitude of the active natural orbital projected into (a) real space and (b) momentum space with a chain of 499 lattice sites. Both figures are symmetric about the origin of the coordinate. The Fermi energy of the tight-binding chain we used is $\varepsilon_{k_F}=0$ with the Fermi wave vector $k_F=\pm \frac{\pi}{2}$, and the dispersion relation is given by $\varepsilon_k=-2t\cos(k)$.}
\end{figure}

\begin{figure}[h!]
\centering
\includegraphics[width=1\columnwidth]{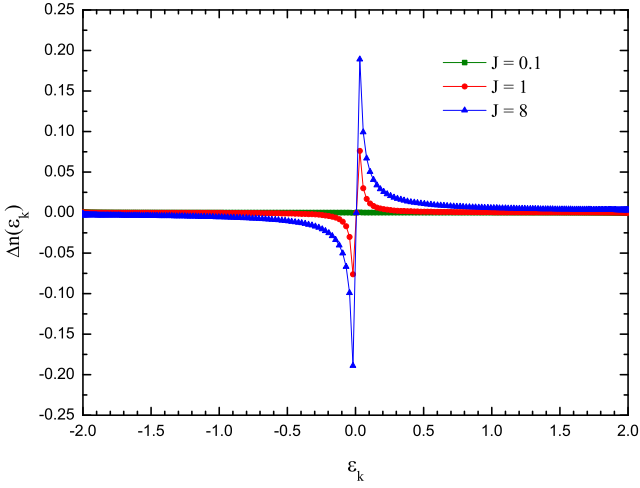}
\caption{\label{fig:DevOcc}(color online) Variations of the electron occupancy number $\Delta n (\varepsilon_k)$ from those for the Kondo coupling $J=0$. Here the system size in calculation is $L = 499$.}
\end{figure}

The amplitude of the active natural orbital projected into real space $|w_i|^2=|U_{{\text{ANo}},i}^ {\dagger}|^2$ and momentum space $|u_k|^2=|\frac{1}{\sqrt{N}}\sum_{l=1}^Ne^{-ikl}U_{{\text{ANo}},l}^ {\dagger}|^2$ are shown in Fig.~\ref{fig:NOAmp}(a) and Fig.~\ref{fig:NOAmp}(b), respectively. In the weak coupling regime $J \to 0$, all the sites namely Wannier orbitals in real space tend to equally compose the active natural orbital. In contrast, in momentum space, single-particle states near the Fermi energy (low-energy excitations) dominantly participate in constituting the active natural orbital, i.e., $|u_{k_F=\pm \pi/2}|^2 \to 0.5$ when $J \to 0$. As the Kondo coupling $J$ increases, the amplitude $|w_0|^2=|U_{{\text{ANo}},0}^ {\dagger}|^2$ projected into the central site (indexed as site 0), which links directly with the magnetic impurity, becomes dominant, while the single-particle states with higher energies come into constituting the active natural orbital. In the strong coupling limit $J \to \infty$, the amplitude $|w_0|^2 \to 1$ and the others decay dramatically with site $i$, meaning the active natural orbital becomes localized $d_{m\sigma }^ {\dagger}=c_{0\sigma }^ {\dagger}$, while the single-particle states with higher energies in momentum space tend to compose the active natural orbital equally. Figure~\ref{fig:DevOcc} presents the variation of electron occupancy number $\Delta n (\varepsilon_k)$. The variation is nearly inversely proportional to the energy nearby the Fermi energy $\varepsilon_{k_F}=0$, and the extent of deviation from the Fermi distribution increases with the Kondo coupling $J$. As we see, the variation is essentially centralized around the Fermi energy. This is actually a manifestation of the Kondo resonance.

\subsection{Orbital-resolved frequency spectra}
\label{sec:Spectral function}
\begin{figure}[h!]
\centering
\includegraphics[width=1\columnwidth]{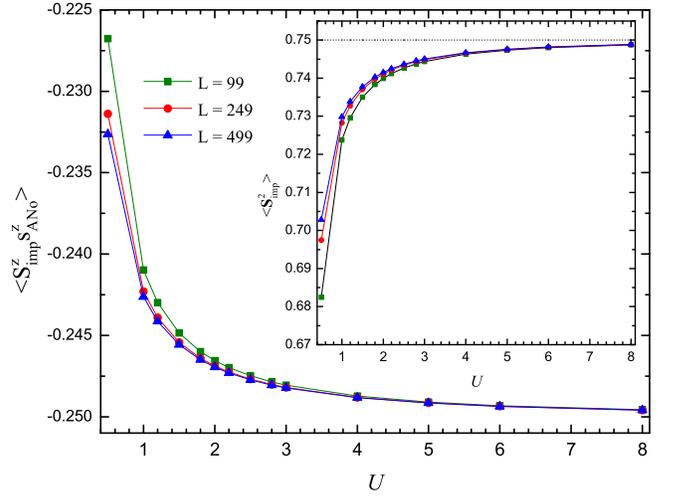}
\caption{\label{fig:SIAMSpinCorr}(color online) Spin-spin correlation function between the active natural orbital and magnetic impurity $\langle {\bf{S}}_{\text{imp}}^z{\bf{s}}_{\text{ANo}}^z\rangle$ as a function of the Hubbard $U$ with chain size $L = 99, 249$ and 499. The inset shows the local moment $\langle{\bf{S}}_{\text{imp}}^2\rangle$ at the impurity site with the hybridization $V=0.2$. The dotted line marks $\langle{\bf{S}}_{\text{imp}}^2\rangle$ in the strong coupling limit $U \to \infty$.}
\end{figure}

Now we come to clarify the orbital-resolved frequency spectra, especially the active natural orbital-resolved, which are represented by the Green's functions. In this subsection we study the single-impurity Anderson model (Eq.~(\ref{eq:Anderson})) on a 1D chain, schematically shown in Fig.~\ref{fig:Model}, which contains the charge fluctuations at the impurity site in comparison with the Kondo model. For the ground state of the single-impurity Anderson model, our calculations show that there also exists only one active natural orbital (numerical results not shown). We then compute the spin-spin correlation function between the active natural orbital and Anderson impurity $\langle {\bf{S}}_{\text{imp}}^z{\bf{s}}_{\text{ANo}}^z\rangle$. The corresponding numerical results with hybridization $V=0.2$ in Eq.~(\ref{eq:Anderson}) are shown in Fig.~\ref{fig:SIAMSpinCorr}, and the inset presents the local moment $\langle{\bf{S}}_{\text{imp}}^2\rangle$ at the impurity site as a function of the Hubbard $U$. As we see, the local moment at the impurity site goes to spin-1/2 as $U$ increases. And we also find that the impurity spin is screened dominantly by the active natural orbital, which is certainly consistent with the above results calculated with the Kondo model.

We now consider the following electronic one-particle Green's function defined at sites or orbitals,
\begin{equation}
\begin{split}
G^\sigma_{ij}(\omega) =& \langle0|c_{i\sigma}\frac{1}{\omega+i\eta-(H-E_0)}c^\dagger_{j\sigma}|0\rangle  \\
                       & +\langle0|c^\dagger_{j\sigma}\frac{1}{\omega+i\eta+(H-E_0)}c_{i\sigma}|0\rangle,
\label{eq:Greenfunction}
\end{split}
\end{equation}
where $|0\rangle$ and $E_0$ mean the ground state and ground-state energy respectively. Equation~(\ref{eq:Greenfunction}) is also commonly used to calculate spectral function within the DMRG \cite{Till D. Kuhner1999, Robert Peters2011,P. E. Dargel2012}. The spectral function is given by $A^\sigma_{ij}(\omega)=-{\text{Im}}(G^\sigma_{ij}(\omega))/\pi$ with the Lorentzian broadening factor $\eta \to 0$. We focus on the single-particle local density of states $\rho(\omega)=-{\text{Im}}(G^\sigma_{ii}(\omega))/\pi$ with a finite $\eta$, for the calculation of which the widely used approach namely the correction vector method\cite{Till D. Kuhner1999,Jeckelmann2002} is adopted. Here the one-particle Green's function defined at the impurity site and at the active natural orbital are computed with $\eta$ being set to 0.02 for the ground state of the Anderson model (Eq.~(\ref{eq:Anderson})) on a chain of $L$ = 249 lattice sites.

\begin{figure}[h!]
\centering
\includegraphics[width=1\columnwidth]{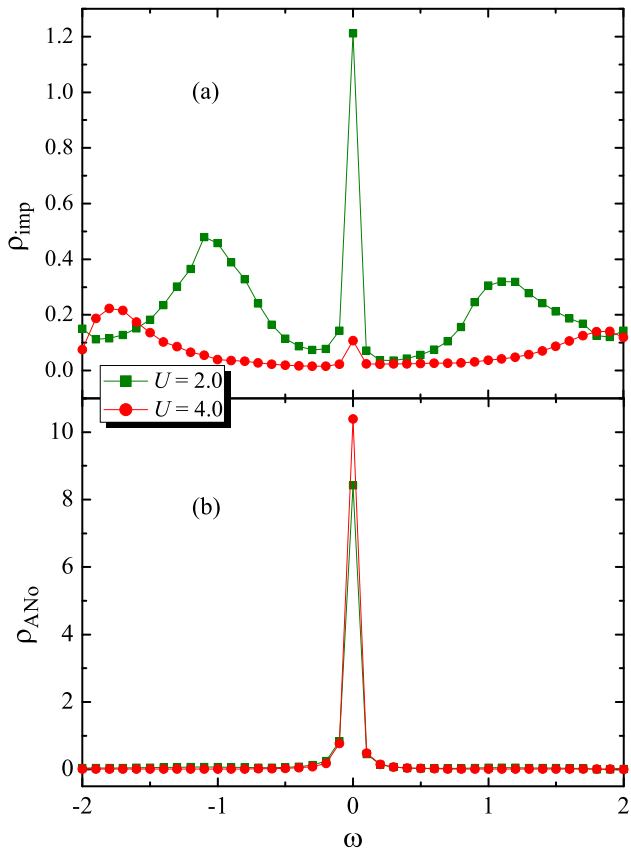}
\caption{\label{fig:DOS}(color online) Single-particle local density of states (a) at the impurity site and (b) at the active natural orbital calculated on a 1D chain with $U=2.0,4.0$, and hybridization $V=0.5$, respectively. The calculation was carried out on a chain of $L = 249$ lattice sites.}
\end{figure}

Figure~\ref{fig:DOS} (a) and (b) show the calculated local density of states at the impurity site and at the active natural orbital, $\rho_{\text{imp}}(\omega)$ and $\rho_{\text{ANo}}(\omega)$, respectively. From Fig.~\ref{fig:DOS} (b), we see the Kondo resonance at $\omega=0$ namely the Fermi energy. The resonance peak is enhanced with increasing Hubbard $U$, consistent with the deviation of  electron occupancy number shown in Fig.~\ref{fig:DevOcc}. On the other hand, at the impurity site, from Fig.~\ref{fig:DOS} (a) we find that there is charge fluctuation corresponding to the Kondo resonance at $\omega=0$. This charge fluctuation is gradually suppressed when increasing the Hubbard $U$. In the limit $U \to \infty $, the local density of states at the impurity site becomes zero namely a local magnetic impurity moment formed without any charge fluctuation at the impurity site.

\subsection{Kondo screening cloud}
\label{sec:Screening cloud}
While there have been many theoretical and experimental investigations\cite{Boyce1974,Madhavan1998,Manoharan2000,Wenderoth2011,Y2007} on the detection of a Kondo screening cloud, it has never been observed experimentally. The Kondo screening cloud manifests itself in the spin-spin correlations between the impurity and the conduction electrons. Accordingly we define a Kondo correlation energy $E_{\text{corr}}(L,J,r)$ to characterize the spatial distribution of a Kondo screening cloud as follows,
\begin{equation}
E_{\text{corr}}(L,J,r)=J|\langle {\bf{S}}_{\text{imp}} \cdot {\bf{s}}(r) \rangle|,
\label{eq:KondoEnergy}
\end{equation}
where $J$ is the Kondo coupling and ${\bf{s}}(r)$ is the spin of conduction electron at a distance $r$ with respect to the impurity site.

\begin{figure}[h!]
\centering
\includegraphics[width=1\columnwidth]{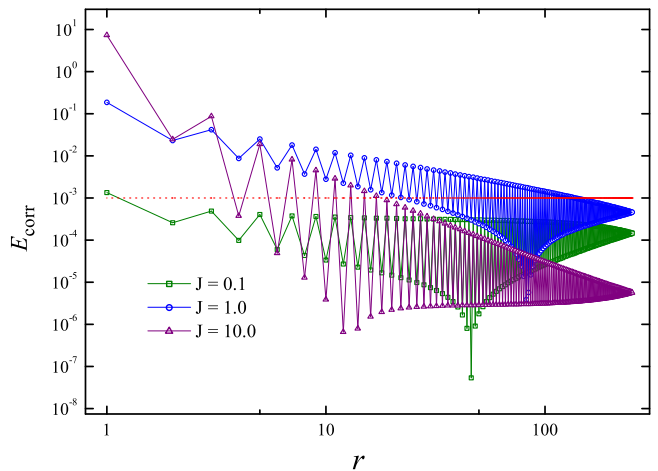}
\caption{\label{fig:KondoEnergy}(color online) Kondo correlation energy calculated on a 1D chain with size $L = 499$ as a function of distance $r$ with respect to the impurity site for $J = 0.1, 1.0,$ and $10.0$ respectively. }
\end{figure}

Figure~\ref{fig:KondoEnergy} presents the Kondo correlation energy calculated on a 1D chain of size $L = 499$ with various Kondo coupling $J$ as a function of distance $r=|i|+1$ from
site $i$ to the magnetic impurity. In the weak coupling regime $J = 0.1$, the envelope of Kondo correlation energy decays very slowly with the distance $r$, but its values are extremely small at all the distances. In contrast, in the
strong coupling regime $J = 10.0$, the envelope of Kondo energy decays dramatically with the distance $r$, meaning the Kondo cloud is very localized.

\begin{figure}[h!]
\centering
\includegraphics[width=1\columnwidth]{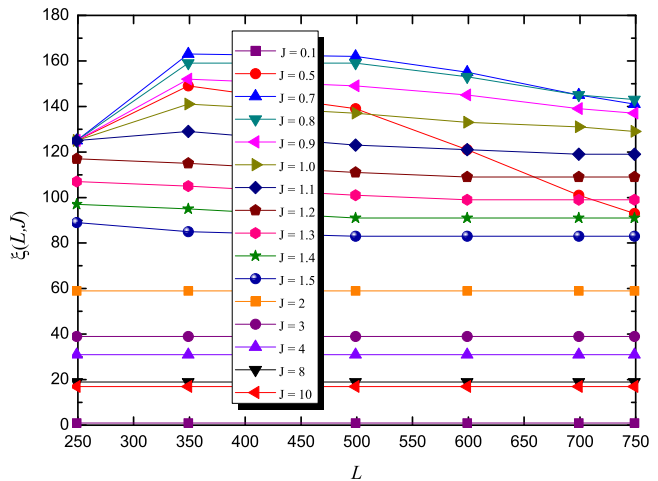}
\caption{\label{fig:Length-L}(color online) Characteristic length $\xi(L,J)$ defined in the text as a function of chain size $L$ for various Kondo couplings $J$. $\xi(L,J)$ reaches its limit value for $J$ in strong coupling regime, however, $\xi(L,J)$ does not tend to saturate for $J$ in weak or intermediate regime when the system size $L$ increases.}
\end{figure}

The resolution limits on energy scale for various spectra measured in experiment are about 1 meV. In reference to the realistic cases, the hopping parameter $t$ in Eq.~(\ref{eq:Hband}) is reasonably set to 1 eV. We thus define the characteristic length scale $\xi(L,J)$ by the value of the maximum of $r$ such that $E_{\text{corr}}(L,J,r)\ge 10^{-3}$, beyond which the Kondo screening cloud cannot be detected in experiment. We can then obtain the value $\xi(L,J)$ for various values of Kondo coupling $J$ and different size $L$, as shown in Fig.~\ref{fig:Length-L}. The length $\xi(J)$ in an infinite host can be extracted, where we
denote $\xi(J)\equiv\xi(\infty, J)$, when we extrapolate $\xi(L,J)$ obtained on chains of different sizes to the thermodynamic limit $L \to \infty$.

\begin{figure}[h!]
\centering
\includegraphics[width=1\columnwidth]{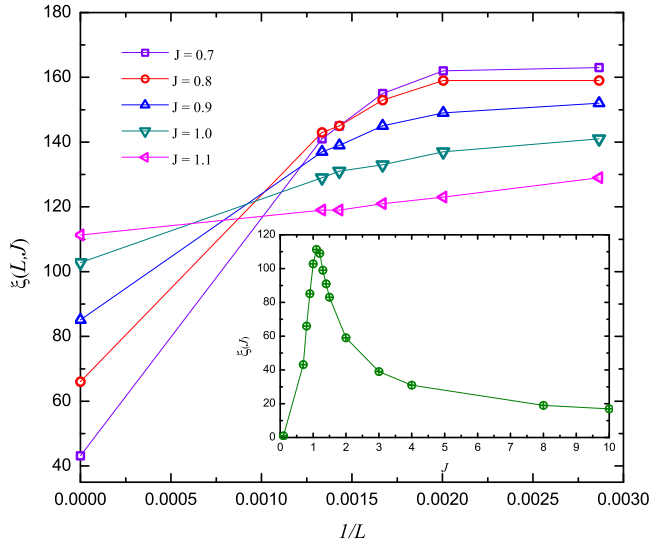}
\caption{\label{fig:Length-J}(color online) The finite size extrapolation of characteristic length $\xi(L,J)$ calculated on a chain. The endpoints at $1/L \to 0$ in the curves are obtained by a quadratic polynomial fit of the five points with $L = 349,499,599,699$ and $749$.}
\end{figure}

Figure~\ref{fig:Length-L} shows the $L$ dependence of $\xi(L,J)$ for various Kondo coupling $J$ with different size $L$ up to $L = 749$. We observe that $\xi(L,J)$ reaches its
limit value for the Kondo couping $J$ in strong coupling regime and then we can get the value of  $\xi(J)$ directly. For the Kondo coupling $J$ in both the weak and the intermediate regimes, however, $\xi(L,J)$ has not saturated for the largest size $L$ we have reached. In this case, the values of $\xi(J)$ are obtained by extrapolating $\xi(L,J)$ to the thermodynamic limit ${1}/{L} \to 0$ using a quadratic polynomial fit, as presented in Fig.~\ref{fig:Length-J}. The derived values of $\xi(J)$ are presented in the inset of Fig.~\ref{fig:Length-J}. We see that the Kondo screening cloud can only be detected at most within 120 lattice spacings (about 20 nm). Actually, in realistic cases, the Kondo coupling $J$ is usually $10\sim100$ meV. The corresponding characteristic length is thus less than 2 nm. This can help us to understand why it is so difficult to detect the Kondo screening cloud experimentally.

\section{Discussion and summary}
\label{sec:summary}
For a long time, the Kondo singlet has been considered as a standard and important many-body state for representing strong correlation and high entanglement. In this work, we study the Kondo singlet by the newly developed natural orbitals renormalization group (NORG) method. We first examine the occupancy of natural orbitals in the ground state of the single-impurity Kondo model. We find that all the natural orbitals rush exponentially into full occupancy or empty except for the single active natural orbital. In both the weak and the strong Kondo coupling regime, this active natural orbital screens the impurity spin solely, forming a spin singlet with the impurity spin, while the subsystem formed by the active natural orbital and the magnetic impurity disentangles from the remaining part of the system. In the intermediate Kondo coupling regime, the magnetic impurity is still screened dominantly by the active natural orbital as well as entangled dominantly with the active natural orbital. Overall, the Kondo singlet can be well approximated as a product state of the spin singlet formed by the magnetic impurity and the active natural orbital with the Fermi sea formed by all the other natural orbitals. The Kondo screening mechanism is thus transparent as well as simple when it is examined in the framework of natural orbitals formalism. This demonstrates that the natural orbitals formalism is an appropriate platform for resolving intrinsic structure of a Kondo singlet. Likewise, we expect that such a picture also works in the Kondo phase of a Kondo or Anderson lattice model, in which there is one active natural orbital screening each Kondo or Anderson site, considering that the number of active natural orbitals is equal to the number of impurities for a multiple-impurity Kondo model \cite{He2014}.

To resolve the structure of a Kondo singlet, we study the structures of the active natural orbital projected into both real space and momentum space. In the weak coupling regime $J \to 0$, all the sites in real space tend to equally compose the active natural orbital. In contrast, in momentum space, single-particle states near the Fermi energy (low-energy excitations) dominantly participate in constituting the active natural orbital. As the Kondo coupling $J$ increases, the site linking directly with the impurity becomes dominant, while the single-particle states with higher energies come into constituting the active natural orbital. In the strong coupling limit $J \to \infty$, the active natural orbital becomes localized, while the single-particle states with higher energies in momentum space tend to compose the active natural orbital equally.

We also study the orbital-resolved frequency spectra of the Kondo singlet by calculating the one-electron Green's function defined at the impurity site and the active natural orbital for the single-impurity Anderson model. The well-known Kondo resonance is clearly shown in the spectral function at the active natural orbital, and its peak is enhanced with increasing Hubbard $U$. On the other hand, at the impurity site, we find a large charge fluctuation corresponding to the Kondo resonance at the Fermi energy. This charge fluctuation is gradually suppressed when increasing the Hubbard $U$. In the limit $U \to \infty $, the local density of states at the impurity site becomes zero namely a local magnetic impurity moment formed without any charge fluctuation at the impurity site.

To clarify the Kondo screening cloud, we introduce the Kondo correlation energy defined by  Eq.~(\ref{eq:KondoEnergy}) in real space to characterize the spatial extension of Kondo screening cloud. We then extract the characteristic length scale within which the Kondo screening cloud can be detected. The calculations reveal that the characteristic length in realistic cases is usually less than 2 nm, which is much smaller than the so-called Kondo length $\xi_k$ estimated as large as 1 $\mu$m in the literature. This explains why the Kondo screening cloud has not been detected experimentally. Our study suggests that an atomic-scale detection tool with very high resolution is needed to detect the Kondo screening cloud in experiment.

In summary, we have investigated the structure of a Kondo singlet by using the newly developed natural orbitals renormalization group. We find that the intrinsic structure of a Kondo singlet can be clearly resolved in the framework of natural orbitals formalism, in which one single active natural orbital plays an essential role. More specifically, the Kondo singlet can be characterized by using the spin singlet formed by the impurity spin and the active natural orbital. It turns out that the Kondo screening mechanism is transparent and simple. Meanwhile we clarify a long standing issue why the Kondo screening cloud has never been detected in experiment.

\begin{acknowledgments}

This work was supported by National Natural Science Foundation of China (Grants No. 11474356 and No. 11774422). Computational resources were provided by National Supercomputer Center in Guangzhou with Tianhe-2 Supercomputer and Physical Laboratory of High Performance Computing in RUC.

\end{acknowledgments}

\end{document}